\documentclass[twocolumn]{aastex631}

\usepackage[utf8]{inputenc}
\usepackage{float}
\usepackage{comment}
\usepackage{rotating}
\usepackage{graphicx}
\usepackage{booktabs}

\shorttitle{The Most Massive White Dwarf Pulsator}
\shortauthors{Caliskan et al.}

\submitjournal{ApJ}
\DeclareUnicodeCharacter{2212}{-}

\begin{document}

\title{Asteroseismology of WD J004917.14-252556.81, the Most Massive Pulsating White Dwarf}

\correspondingauthor{Ozcan Caliskan}
\email{caliskanozcan@hotmail.com}  

\author[0009-0003-5839-8007]{Ozcan Caliskan}
\affiliation{Programme of Astronomy and Space Sciences, Institute of Graduate Studies in Science, Istanbul University, 34116, Istanbul, T\"{u}rkiye}
\affiliation{Homer L. Dodge Department of Physics and Astronomy, University of Oklahoma, 440 W. Brooks Str., Norman, OK, 73019, USA}

\author[0000-0002-6603-994X]{Murat Uzundag}
\affiliation{Institute of Astronomy, KU Leuven, Celestijnenlaan 200D, 3001, Leuven, Belgium}

\author[0000-0001-6098-2235]{Mukremin Kilic}
\affiliation{Homer L. Dodge Department of Physics and Astronomy, University of Oklahoma, 440 W. Brooks Str., Norman, OK, 73019, USA}

\author[0000-0002-8079-0772]{Francisco C. De Gerónimo}
\affiliation{Grupo de Evolución Estelar y Pulsaciones. Facultad de Ciencias Astronómicas y Geofísicas, Universidad Nacional de La Plata, Paseo del
Bosque s/n, (1900) La Plata, Argentina}
\affiliation{Instituto de Astrofísica de La Plata, IALP (CCT La Plata), CONICET-UNLP, Argentina}

\author[0000-0001-7143-0890]{Adam Moss}
\affiliation{Homer L. Dodge Department of Physics and Astronomy, University of Oklahoma, 440 W. Brooks Str., Norman, OK, 73019, USA} 

\author[0000-0002-0006-9900]{Alejandro H. Córsico}
\affiliation{Grupo de Evolución Estelar y Pulsaciones. Facultad de Ciencias Astronómicas y Geofísicas, Universidad Nacional de La Plata, Paseo del
Bosque s/n, (1900) La Plata, Argentina}
\affiliation{Instituto de Astrofísica de La Plata, IALP (CCT La Plata), CONICET-UNLP, Argentina}

\author[0000-0002-2695-2654]{Steven G. Parsons}
\affiliation{Astrophysics Research Cluster, School of Mathematical and Physical Sciences, University of Sheffield, Sheffield S3 7RH, UK}

\author[0000-0003-4615-6556]{Ingrid Pelisoli}
\affiliation{Department of Physics, University of Warwick, Gibbet Hill Road, Coventry, CV4 7AL, UK}

\author[0009-0009-9105-7865]{Gracyn Jewett}
\affiliation{Homer L. Dodge Department of Physics and Astronomy, University of Oklahoma, 440 W. Brooks Str., Norman, OK, 73019, USA}

\author[0000-0002-6153-7173]{Alberto Rebassa-Mansergas}
\affiliation{Departament de Física, Universitat Politècnica de Catalunya, c/Esteve Terrades 5, 08860 Castelldefels, Spain}
\affiliation{Institut d'Estudis Espacials de Catalunya, Esteve Terradas, 1, Edifici RDIT, Campus PMT-UPC, 08860 Castelldefels, Barcelona, Spain}

\author[0000-0002-3316-7240]{Alex J. Brown}
\affiliation{Hamburger Sternwarte, University of Hamburg, Gojenbergsweg 112, 21029 Hamburg, Germany}

\author[0000-0003-4236-9642]{Vikram S. Dhillon}
\affiliation{Astrophysics Research Cluster, School of Mathematical and Physical Sciences, University of Sheffield, Sheffield S3 7RH, UK}
\affiliation{Instituto de Astrof\'{i}sica de Canarias, E-38205 La Laguna, Tenerife, Spain}

\author[0000-0003-2368-345X]{Pierre Bergeron}
\affiliation{D\'epartement de Physique, Universit\'e de Montr\'eal, C.P. 6128, Succ. Centre-Ville, Montr\'eal, QC H3C 3J7, Canada}

\begin{abstract}

We present extensive follow-up time-series photometry of WD J0049$-$2525, the most massive pulsating white dwarf currently known with $T_{\rm eff} = 13\, 020\,{\rm K}$ and $\log{\it g} = 9.34$ cm s$^{-2}$. The discovery observations detected only two significant pulsation modes. Here, we report the detection of 13 significant pulsation modes ranging from 170 to 258 s based on 11 nights of observations with the New Technology Telescope, Gemini, and Apache Point Observatory telescopes. We use these 13 modes to perform asteroseismology and find that the best-fitting models (under the assumption of an ONe core composition) have $M_{\star} \approx 1.29~M_\odot$, surface hydrogen layer mass of $\log(M_{\rm H}/M_{\star}) \lesssim -7.5$, and a crystallized core fraction of $>99\%$. An analysis of the period spacing also strongly suggests a very high mass. The asteroseismic distance derived is in good agreement with the distance provided by {\it Gaia}. We also find tentative evidence of a rotation period of 0.3 or 0.67 d.  This analysis provides the first look at the interior of a $\sim 1.3~M_{\odot}$ white dwarf. 

\end{abstract}

\section{Introduction}
\label{sec:intro}

White dwarfs (WDs) represent the final evolutionary stage for the vast majority of stars, specifically low- and intermediate-mass stars that comprise over 95\% of all stars in the Milky Way \citep{2010A&ARv..18..471A,2022PhR...988....1S}.  
During their evolution, WDs pass through at least one phase of pulsational instability, transforming them into pulsating variables.
ZZ Ceti variables are pulsating DA (H-rich atmosphere) WDs with effective temperature ($T_{\rm eff}$) in the range of $10\,500$ K to $13\,500$ K. 
They exhibit pulsation periods between $\sim$100 and $\sim$1400 s due to non-radial gravity ($g$) modes with harmonic degrees ($\ell$) 1 and 2 \citep{Fontaine08, Winget08, 2019A&ARv..27....7C}. Asteroseismology enables us to probe the interiors of these dense objects by comparing theoretically calculated pulsation periods with observations \citep[see, e.g.,][]{1998ApJS..116..307B,Córsico19}.

Ultra-massive DA WDs, which are characterized by $M_{\star} \gtrsim 1.05~M_{\odot}$,  are expected to have cores composed of $^{16}$O and $^{20}$Ne \citep{Schwab21}. This is because their progenitor stars burnt semi-degenerate carbon during their evolution on the super-asymptotic giant branch phase. However, \cite{2021A&A...646A..30A} suggest that single star evolution could lead to ultra-massive WDs with $^{12}$C and $^{16}$O cores. Binary mergers may also produce ultramassive CO core WDs \citep[like DAQ WDs,][]{Hollands20,Shen23,Kilic24,Jewett24}. There is growing evidence that a significant fraction of ultra-massive WDs in the solar neighborhood suffer from $^{22}$Ne distillation that only occurs in CO-core WDs \citep{Cheng19,Blouin21,Bedard24,Kilic25}.

By the time ultramassive WDs reach the ZZ Ceti instability strip ($T_{\rm eff} \sim 12\, 500\,{\rm K}$), crystallization is predicted to occur. Theoretical models suggest that crystallization leads to a separation of $^{16}$O and $^{20}$Ne (or $^{12}$C and $^{16}$O) in the interior of ultra-massive WDs, which affects their pulsational properties. This characteristic offers a unique opportunity to study the crystallization processes \citep{2019A&A...621A.100D,Córsico20}. Detection of a significant number of pulsation modes in the ultra-massive H-atmosphere WD BPM\,37093 \citep{Kanaan92} provided the first opportunity to search for the observational hallmark of crystallization in a single WD star. Depending on its mass and internal composition, the core of this star may be up to 90\% crystalline \citep[][though also see \citealt{Brassard05}]{Winget97,Montgomery99,Metcalfe04,Córsico19}.

ZZ Ceti stars can be subdivided into three subclasses: hot, intermediate, and cold, based on their effective temperature and pulsation characteristics \citep{Clemens94, Mukadam06}. Hot ZZ Cetis are located at the blue edge of the instability strip and exhibit stable sinusoidal or jagged light curves, with a few modes having short periods ($\lesssim 350$~s) and small amplitudes ($1.5–20$ mma). Cool ZZ Cetis, on the other hand, are located at the red edge of the instability strip, showing a collection of long periods (up to $\sim$1500 s) and large variation amplitudes ($40–110$ mma). Their light curves are non-sinusoidal and suffer from significant mode interference. Finally, the intermediate ZZ Ceti stars show mixed characteristics from both hot and cool members \citep{Romero22}.

ZZ Ceti stars typically have masses in the range of $0.5\ M_{\odot} < M_{\star} < 0.8\ M_{\odot}$. However, eight ultra-massive ZZ Ceti stars with $M\geq1.05~M_{\odot}$ have been discovered so far: BPM~37093 \citep[$M_{\star} = 1.13~M_{\odot}$,][]{Kanaan92, Bedard17}, GD~518 \citep[$M_{\star} = 1.24~M_{\odot}$,][]{Hermes13}, SDSS~J084021.23+522217.4 \citep[$M_{\star}= 1.16 M_{\odot}$,][]{Curd17}, WD J212402.03$-$600100.05 \citep[$M_{\star} = 1.16~M_{\odot}$,][]{Rowan19}, WD J0204+8713 \citep[$M_{\star} = 1.05~M_{\odot}$,][]{Vincent20,Jewett24}, WD J0551+4135 \citep[$M_{\star}= 1.13~M_{\odot}$,][]{Vincent20,Hollands20}, WD J004917.14$-$252556.81 \citep[$M_{\star} \sim 1.30 M_{\odot}$,][]{Kilic23}, and finally WD J0135+5722 \citep[$M_{\star} = 1.12 - 1.15~M_{\odot}$,][]{degeranimo25}.  With such a high mass, the ultra-massive ZZ Ceti star WD J0049$-$2525 is the most massive pulsating WD currently known.

\begin{table}[t]
\setlength{\tabcolsep}{1pt}
\centering
\caption{Observational and physical properties of WD J0049$-$2525.} 
\begin{tabular}{|l c|}
\hline
\multicolumn{2}{|c|}{GAIA DR3 Parameters} \\
\hline
ID & 2345323551189913600 \\
RA (h:m:s) & ~~00 49 17.14 \\
DEC (d:m:s) & $-$25 25 56.81 \\
$G$ (mag) & 19.04 \\
$G_{\rm BP}$ (mag) & 19.08 \\
$G_{\rm RP}$ (mag) & 19.05 \\
$\varpi$ (mas) & 10.04 \\
$d$ (pc) \citep{Bailer-Jones21} & $99.7^{+2.9}_{-2.7}$ \\
$\mu_{\alpha}\cos\delta$ (mas yr$^{-1}$) & 22.54 \\
$\mu_{\delta}$ (mas yr$^{-1}$) & $-$28.35 \\
RUWE & 1.054 \\
\hline
\multicolumn{2}{|c|}{Physical Properties \citep{kilic23merger}} \\
\hline
$T_{\rm eff}$ (K) & $13\,020\pm460$ \\
$\log g$ (cm s$^{-2}$) & $9.34\pm0.04$ \\
Mass, ONe Core ($M_{\odot}$) & $1.26\pm0.01$ \\
Cooling Age, ONe Core (Gyr) & $1.94\pm0.08$\\
Mass, CO Core ($M_{\odot}$) & $1.31\pm0.01$ \\
Cooling Age, CO Core (Gyr) & $1.72\pm0.09$\\
\hline
\end{tabular}
\label{tab:Props}
\end{table}

The discovery and characterization of pulsating ultra-massive WDs through asteroseismology is crucial for our understanding of their interior structures and their relation with Type Ia supernovae \citep[e.g.,][]{Nugent11,Maoz14}. Asteroseismology of the few ultra-massive WDs currently known can provide a unique opportunity to probe their interiors, with potential constraints on their core composition \citep[e.g.,][]{2019A&ARv..27....7C}. 

\begin{figure}
    \centering
    \includegraphics[width=2.5in]{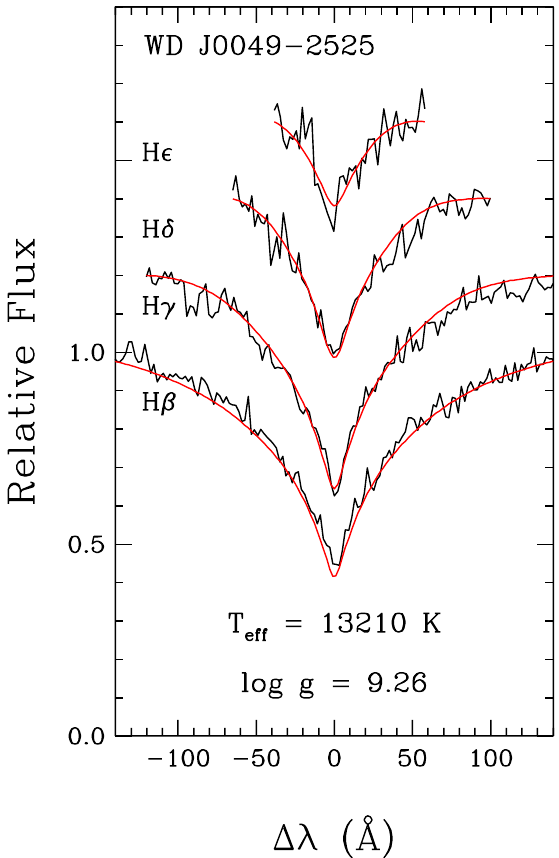}
    \caption{Model fits (red lines) to the normalized Balmer line profiles (black lines) of WD J0049$-$2525. 3D hydrodynamical corrections are included in the best-fitting model parameters.}
    \label{fig:spec}
\end{figure}

In this paper, we focus on the most massive pulsating WD known to date, WD J0049$-$2525. Table \ref{tab:Props} presents the observational and physical parameters of WD J0049$-$2525. \citet{kilic23merger} obtained an optical spectrum of this object, which
confirmed it as an ultramassive DA WD with $T_{\rm eff} = 13\,020\pm460$ K and $\log g = 9.34\pm0.04$ cm s$^{-2}$ based on the photometric method \citep{bergeron19}. 
Figure \ref{fig:spec} presents our best fits to the normalized Balmer line profiles using 1D model atmospheres. Including the 3D hydrodynamical  corrections from \citet{Tremblay13}, the best-fitting parameters are $T_{\rm eff} = 13\,210\pm360$ K and $\log g = 9.26\pm0.05$ cm s$^{-2}$. These are consistent with the results from the photometric method (using {\it Gaia} parallax and Pan-STARRS $grizy$ photometry) within the errors, providing further evidence of a very high mass.

\citet{Kilic23} discovered photometric variations in this star based on four nights of observations with total baselines of $\leq2$ h on each night. However, with only two pulsation modes detected in those light curves, their asteroseismic analysis was limited and they could not obtain robust constraints. Here, we present extended follow-up observations of WD J0049$-$2525, and significantly improve mode detection in this star. We describe our observations in Section \ref{obs}, and present the light curves and
frequency analysis in Section \ref{analysis}. We discuss the results from our detailed asteroseismic modeling in Section \ref{model}, and discuss and conclude our findings in Section \ref{conclusion}.

\newpage
\section{Observations}
\label{obs}

We obtained time-series photometry of WD J0049$-$2525 at three different facilities over 7 different nights between 2023 October 5 and 2024 October 20. Table \ref{tab:ObsLogs} presents the observation log for this study.

At the 3.5-m New Technology Telescope (NTT) at La Silla, we used the high-speed camera ULTRACAM \citep{Dhillon07}. ULTRACAM uses a triple beam setup and three frame transfer CCD cameras, which allows simultaneous data in three different wavebands with negligible (24 ms) dead-time between exposures. For our observations we used the high throughput super-SDSS $u$, $g$, and $r$ filters with exposure times of 20,  7, and 7 s, respectively. We obtained simultaneous $ugr$ photometry of WD J0049$-$2525 over the entire nights for 2023 Oct 5, 6, and 7. In total we obtained 4942, 14810, and 15011 $u$, $g$, and $r$-band images with ULTRACAM, respectively \citep{Dhillon21}.

For Apache Point Obervatory (APO) 3.5 telescope run on 2024 January 7, we used the Astrophysical Research Consortium Telescope Imaging Camera (ARCTIC). To reduce the read-out time, we used the quad amplifier mode and binned the CCD by $3\times3$, which resulted in a plate scale of 0.342 arcsec pixel$^{−1}$ and a read-out time of 4.5 s. With 20 s exposures, this resulted in a cadence of 24.5 s.

At the 8m Gemini South telescope, We obtained time-series photometry on 2024 July 16, September 28, and October 20 as part of the program GS-2024B-Q-304. We obtained 293, 293, and 231 back-to-back exposures over those three nights, respectively. To reduce the read-out time, we binned the chip by 4×4, which resulted in a plate scale of 0.32 arcsec pixel$^{−1}$ and a 15.7 s overhead, resulting in a cadence of 22.7 s with our 7 s long exposures.

In addition, we take advantage of the photometry data presented in \citet{Kilic23}, which include two nights of observations each at
APO 3.5m and Gemini South 8m telescopes obtained over the period 2022 December 22 to 2023 January 8. Hence, our final dataset includes
frequency measurements from 11 nights in total.

\begin{figure*}[htp]
    \centering
    \includegraphics[width=0.9\textwidth]{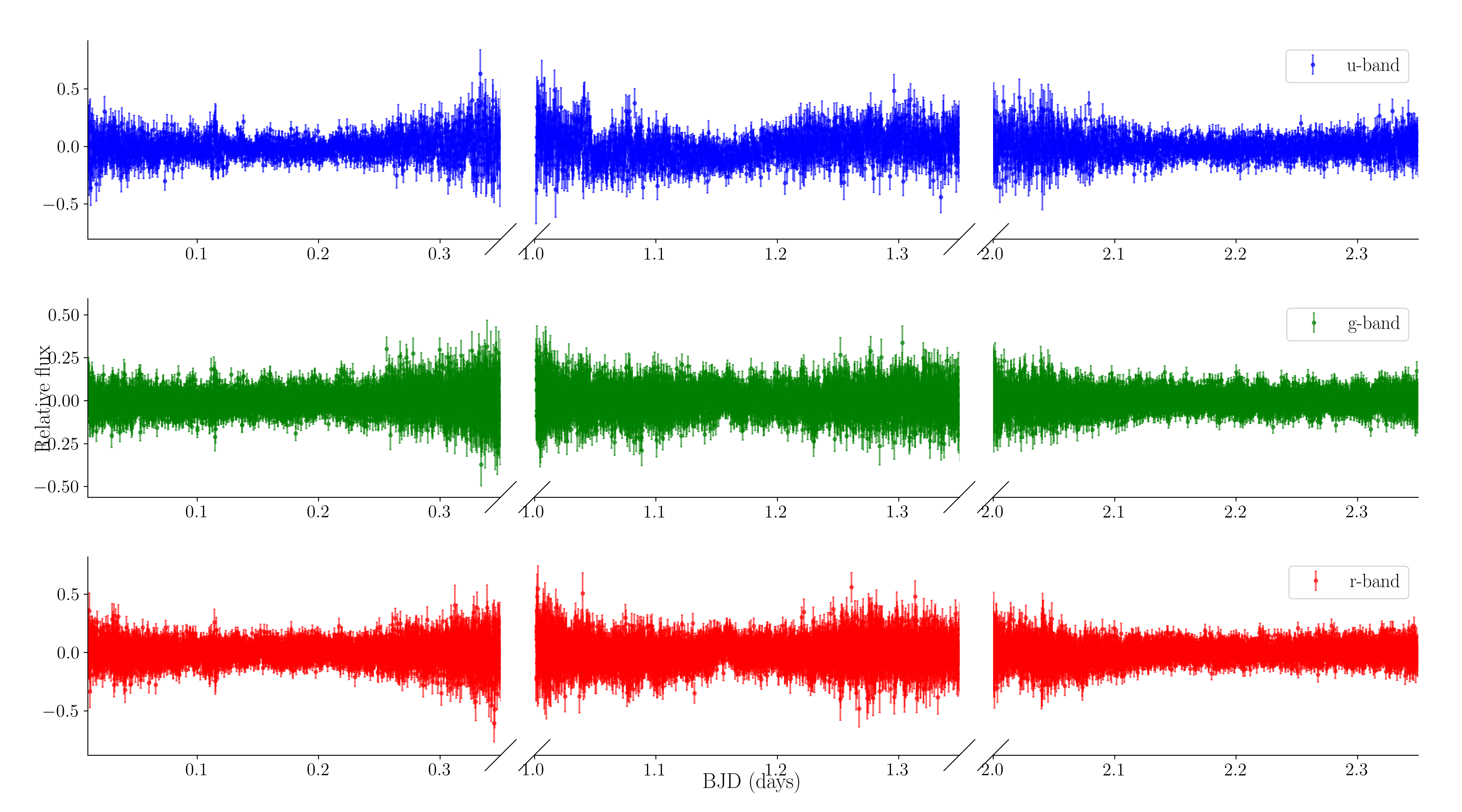}
    \caption{ULTRACAM light curves of WD J0049$-$2525 from UT 2023 October 5, 6, and 7 (left to right). 
    The top, middle, and bottom panels show the relative flux variations in the $u$ (blue), $g$ (green), and $r$ (red) bands, respectively.}
    \label{fig:Ultracam_LC}
\end{figure*}

\begin{figure*}[htp]
    \centering
    \includegraphics[width=0.9\textwidth]{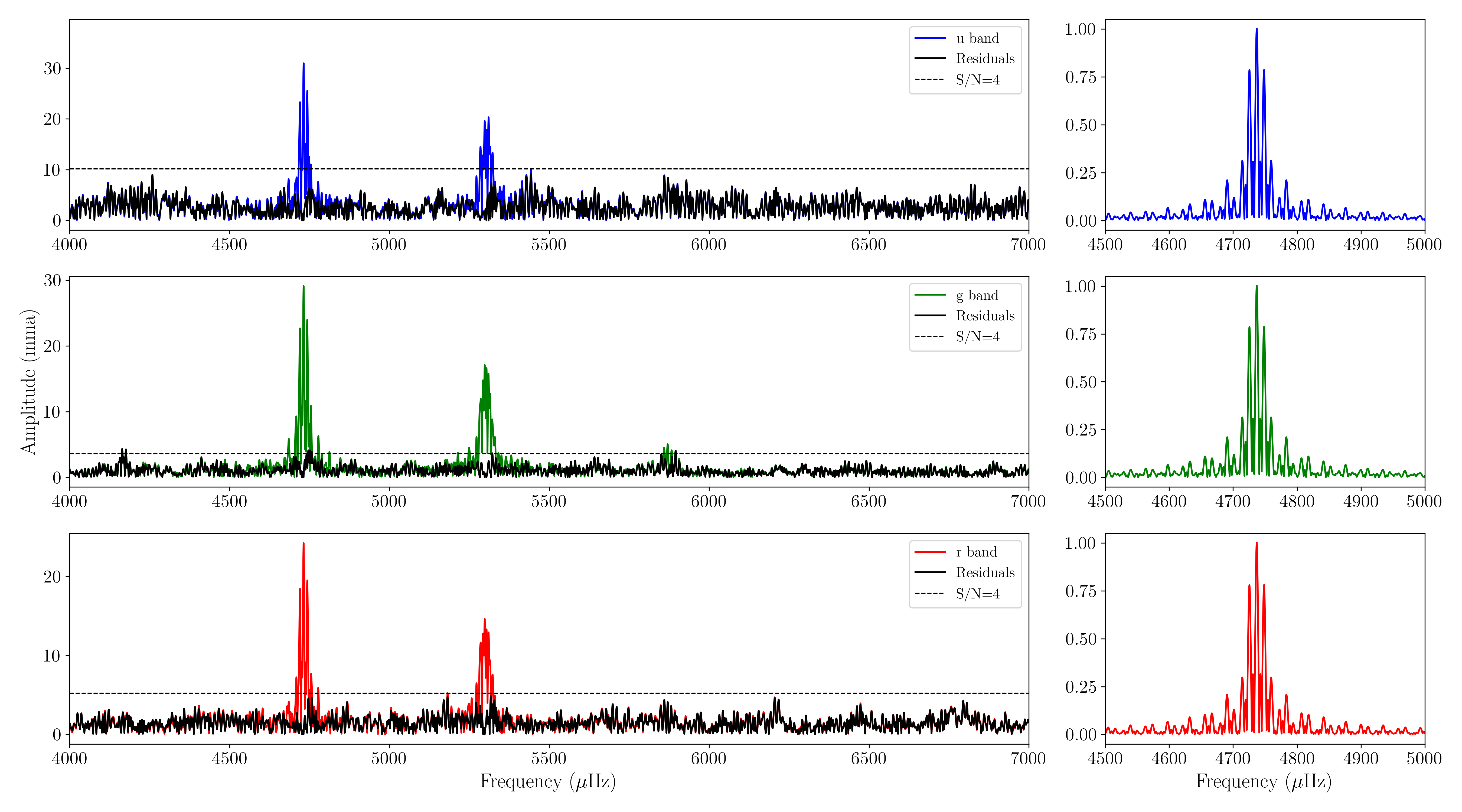}
    \caption{Fourier transforms of the combined ULTRACAM dataset (shown in Figure \ref{fig:Ultracam_LC}) on WD J0049$-$2525. The $u$,
    $g$, and $r$-band data are shown in the top, middle, and bottom panels, respectively. The dashed horizontal line in each panel indicates the 4×$S/N$ detection threshold. The black line is the Fourier transform obtained after the prewhitening procedure discussed in the text. 
    The right panels display the spectral window for each light curve centered at 4737 $\mu$Hz for a comparison.}
    \label{fig:Ultracam_FT}
\end{figure*}

\begin{table}
\setlength{\tabcolsep}{3pt}
  \centering
\caption{Observation log for WD J0049$-$2525. Exposure times for NTT represent $ugr$ bands respectively.}
\begin{tabular}{cccc}
\hline
\hline
UT Date & Instrument & Length of Obs. & Exp. time \\
(yyyy.mm.dd) &  & (h) & (s) \\
\hline
2023-10-05 & NTT       & 8.7 & 20/7/7 \\
2023-10-06 & NTT       & 9.1 & 20/7/7 \\
2023-10-07 & NTT       & 9.2 & 20/7/7 \\
2024-01-07 & APO 3.5 m & 1.9 & 20 \\
2024-07-16 & Gemini    & 1.9 & 7 \\
2024-09-28 & Gemini    & 1.9 & 7 \\
2024-10-20 & Gemini    & 1.5 & 7 \\
\hline
\hline
\end{tabular}%
\label{tab:ObsLogs}%
\end{table}%

\section{Analysis}
\label{analysis}

\begin{table*}[ht]
\setlength{\tabcolsep}{8pt}
\centering
\caption{Pulsation frequencies for WD J0049$-$2525 used in the asteroseismic fit. Rows with an asterisk (*) were calculated using the observations in \citet{Kilic23}.}
\begin{tabular}{clcrrcl}
\hline
\hline
ID & \multicolumn{1}{c}{Frequency} & Period & \multicolumn{1}{c}{Amplitude} & $S/N$ & Instrument & \multicolumn{1}{c}{Date} \\
            & \multicolumn{1}{c}{($\mu$Hz)} & (s) & \multicolumn{1}{c}{(mma)} & & & \multicolumn{1}{c}{(yyyy-mm-dd)} \\
\hline
F1 & $3868.4 \pm 2.3$ & $258.5 \pm 0.2$ & $36.2 \pm 1.0$ & $25.85$ & Gemini & 2024-09-28 \\

F2 & $4160.7 \pm 6.3$ & $240.3 \pm 0.4$ & $27.2 \pm 2.1$ & $10.86$ & APO     & 2024-01-07 \\

F3 & $4354.0 \pm 7.3$ & $229.7 \pm 0.4$ & $21.9 \pm 1.6$ & $10.96$ & Gemini  & 2024-10-20 \\

F4 & $4513.3 \pm 5.8$ & $221.6 \pm 0.3$ & $25.6 \pm 1.3$ & $15.05$ & Gemini  & 2022-12-27* \\

F5 & $4533.0 \pm 8.3$ & $220.6 \pm 0.4$ & $28.0 \pm 3.0$ & $7.36$  & APO     & 2022-12-22* \\

F6 & $4565.3 \pm 15.1$& $219.0 \pm 0.7$ & $11.2 \pm 2.1$ & $4.49$  & APO     & 2024-01-07 \\

F7 & $4644.7 \pm 0.4$ & $215.3 \pm 0.1$ & $40.0 \pm 1.4$ & $20.00$ & Gemini  & 2022-12-26* \\

F8 & $4709.1 \pm 3.0$ & $212.4 \pm 0.1$ & $50.1 \pm 1.8$ & $22.79$ & Gemini  & 2024-07-16 \\

F9 & $4731.7 \pm 0.2$ & $211.3 \pm 0.1$ & $31.3 \pm 2.3$ & $12.33$ & ULTRACAM& 2023-10-05 \\

F10& $4775.7 \pm 10.3$& $209.4 \pm 0.5$ & $14.3 \pm 1.3$ & $8.42$  & Gemini  & 2022-12-27* \\

F11& $4861.6 \pm 9.8$ & $205.7 \pm 0.4$ & $15.4 \pm 1.8$ & $6.99$  & Gemini  & 2024-07-16 \\

F12& $5297.7 \pm 0.1$ & $188.8 \pm 0.1$ & $17.7 \pm 0.8$ & $19.68$ & ULTRACAM& 2023-10-06 \\

F13& $5870.1 \pm 0.4$ & $170.4 \pm 0.1$ & $5.2 \pm 0.8$  & $5.81$  & ULTRACAM& 2023-10-06 \\

\hline
\hline
\end{tabular}
\label{tab:Frequencies}
\end{table*}

\subsection{Light Curve Analysis}

We employed a two-step approach to analyze the light curves obtained from the APO 3.5 m telescope, Gemini, and NTT/ULTRACAM observations, with the primary goal of improving the precision of the frequency analysis. Initially, a five standard deviations (5$\sigma$) clipping method was utilized to detect and eliminate outliers in the data. This process involves calculating the mean and standard deviation of the flux values, then discarding data points that deviate from the mean by more than 5$\sigma$. After this clipping procedure, we applied a second-order polynomial detrending technique to eliminate any long-term systematic variations, which could mask the periodic signals of interest in the light curves. This step involved fitting a polynomial curve to the clipped data and subtracting the fitted curve from the original dataset, isolating the short-term fluctuations. 
The processed light curves for APO and Gemini observations are presented in the Appendix \ref{sec:Appendix}, while ULTRACAM observations are shown in Figure \ref{fig:Ultracam_LC}.

\subsection{Frequency Analysis}

To identify the periodicities within the light curves of WD J0049$-$2525, we performed a Fourier Transform (FT) analysis of each light curve. This allowed us to identify the pulsation frequencies of WD J0049$-$2525, along with their respective amplitudes, phases, and associated errors.

The FTs of our observations spanning seven nights are presented in Figures \ref{fig:Ultracam_FT} and \ref{fig:APO-GEMINI} (right panels). Figure \ref{fig:Ultracam_FT} displays the FTs of the ULTRACAM observations, while Figure \ref{fig:APO-GEMINI} presents those obtained from the APO and Gemini data. We also include four nights of observational data—two nights with APO and two with the Gemini Observatory—that were both acquired and previously used by Kilic et al. (2023), and are now incorporated into our FT analysis.
 
For each FT, we calculated the median noise level and established a detection threshold corresponding to a signal-to-noise ratio (S/N) of 4. This threshold, marked by dashed lines in Figures~\ref{fig:Ultracam_FT} and \ref{fig:APO-GEMINI}, follows the standard approach commonly adopted for ground-based photometry \citep[e.g.,][]{Sowicka23, degeranimo25}.
We then applied a non-linear least squares (NLLS) fitting procedure using the software {\tt Period04} \citep{Lenz05}. This iterative approach involves identifying the most significant peak above the threshold, fitting it, subtracting the corresponding sinusoidal signal from the data, and repeating the process until no additional peaks exceed the detection threshold within the frequency resolution of the dataset.
Each peak was modeled with a sinusoidal function of the form $A_i \sin(\omega_i\ t + \phi_i)$ with $\omega=2\pi/P$, where $A$ is the
amplitude, $P$ is the period, and $\phi$ is the phase. This allowed us to accurately determine the frequency (or period), amplitude, and phase associated with each detected signal.

To ensure a robust identification of pulsation frequencies, we applied two complementary approaches to the ULTRACAM dataset. First, we analyzed each night individually to capture any potential amplitude or phase variability across nights. 

In total, the nightly analysis yielded 24 detected frequencies, mostly concentrated between 4730 $\mu$Hz (211.36 s) and 5312 $\mu$Hz (188.27 s), with strong pulsations evident in all bands. Notable examples include 4513.3539 $\mu$Hz (221.56 s) with $S/N$ = 15.05, 4728.7463 $\mu$Hz (211.47 s) with $S/N$ = 20.81, and 5294.3727 $\mu$Hz (188.94 s) with $S/N$ = 14.62. Three possible combination frequencies were also identified: 10022 $\mu$Hz (4728 + 5294), 575 $\mu$Hz (likely a difference between 5880 and 5310), and 9480 $\mu$Hz.

To complement the nightly analysis, we combined the three ULTRACAM nights into a single dataset to enhance frequency resolution, improve signal-to-noise, and reduce aliasing effects.
In Figure \ref{fig:Ultracam_LC}, we present the combined light curves in the $u$, $g$, and $r$ bands from top to bottom, respectively. Figure \ref{fig:Ultracam_FT} shows the corresponding FTs of these light curves in the same order. For comparison, each panel also includes the spectral window function, centered at 4737~$\mu$Hz, to illustrate the effect of the temporal sampling on the frequency spectrum.
The combined analysis confirms the main pulsation modes found in individual nights and helps refine the frequency and amplitude estimates. Most prominently, the 171 s, 188 s, and 211 s modes were consistently detected in both approaches, reinforcing their significance for asteroseismic modelling. Although the combined dataset complicates the window function and may mask time-dependent amplitude or phase variations, it provides a more complete picture of the star’s dominant pulsation spectrum. We therefore adopt a hybrid strategy, using both analyses to define a robust set of modes.

Gemini observations, including data from three nights obtained in this work and two additional nights previously published by \citet{Kilic23}, span five distinct epochs between December 2022 and October 2024, and contribute a total of 20 pulsation frequencies ranging from 3854.71 $\mu$Hz (259.39 s) to 9458.98 $\mu$Hz (105.78 s). This subset includes the highest $S/N$ detection at 3868.4033 $\mu$Hz (258.50 s), with $S/N$ = 25.85. Additional strong signals appear within the 210–260 s period range, including frequencies at 3854.7127 $\mu$Hz (259.22 s), with $S/N$ = 22.31, 4353.9702 $\mu$Hz (229.64 s), with $S/N$ = 10.96, and 4644.7343 $\mu$Hz (215.24 s), with $S/N$ = 20. Gemini observations provide a broad variety of pulsation modes, some with exceptionally high $S/N$ values. In addition, there are several combination frequencies with low $S/N$ beyond 7000 $\mu$Hz, such as 7729.2645 $\mu$Hz (129.40 s) with $S/N$ = 4.29 and 7751.8519 $\mu$Hz (129.02 s) with $S/N$ = 6.77.

APO observations, collected over three nights in December 2022, January 2023, and January 2024, reveal four frequencies, primarily clustered between 4533.35~$\mu$Hz (220.6s) and 4735.29$\mu$Hz (211.2s). Two of these observing nights (December 2022 and January 2023) were previously presented by \citet{Kilic23}, while the January 2024 data were obtained as part of this work. These detections include a prominent signal at 4160.7 $\mu$Hz (240.3 s), with $S/N$ = 10.86, observed on 2024 January 7. Other detections include three more peaks located at 4533 $\mu$Hz (220.6 s), 4565 $\mu$Hz (219 s) and 4735 $\mu$Hz (211 s).

Several of these detections coincide with signals previously reported in the literature. Notably, frequencies near $221.42 \pm 0.32 s$  and $209.63 \pm 0.55 s$ were also observed by \citet{Kilic23} in their December 2022 Gemini run, while other modes, including $222.48 \pm 0.38 s$ and $206.52 \pm 0.59 s$, are consistent within uncertainties. Additionally, frequencies near $220.61 \pm 0.40 s$ and $211.18 \pm 0.61 s$ detected at APO on December 22, 2022 and January 8, 2023, respectively, are also present in our dataset, providing strong independent confirmation.

A summary of all pre-whitened frequencies for WD J0049$-$2525, including all datasets, is provided in Table \ref{tab:Full_FT}.
The final combined list of frequencies and their corresponding periods from the Gemini, ULTRACAM, and APO datasets includes 13 significant pulsation peaks shown in Table \ref{tab:Frequencies}. These frequencies range from 3868 $\mu$Hz (258 s) to 5870 $\mu$Hz (170 s). 
We note that the list of 13 pulsation frequencies presented here was obtained by identifying the most robust signals across multiple nights and instruments. While a larger number of significant peaks were initially identified, many of them either appear only in individual nights or fall below the significance threshold in the combined dataset.

\begin{figure*}[htp]
   \centering
   \includegraphics[width=1.0\textwidth]{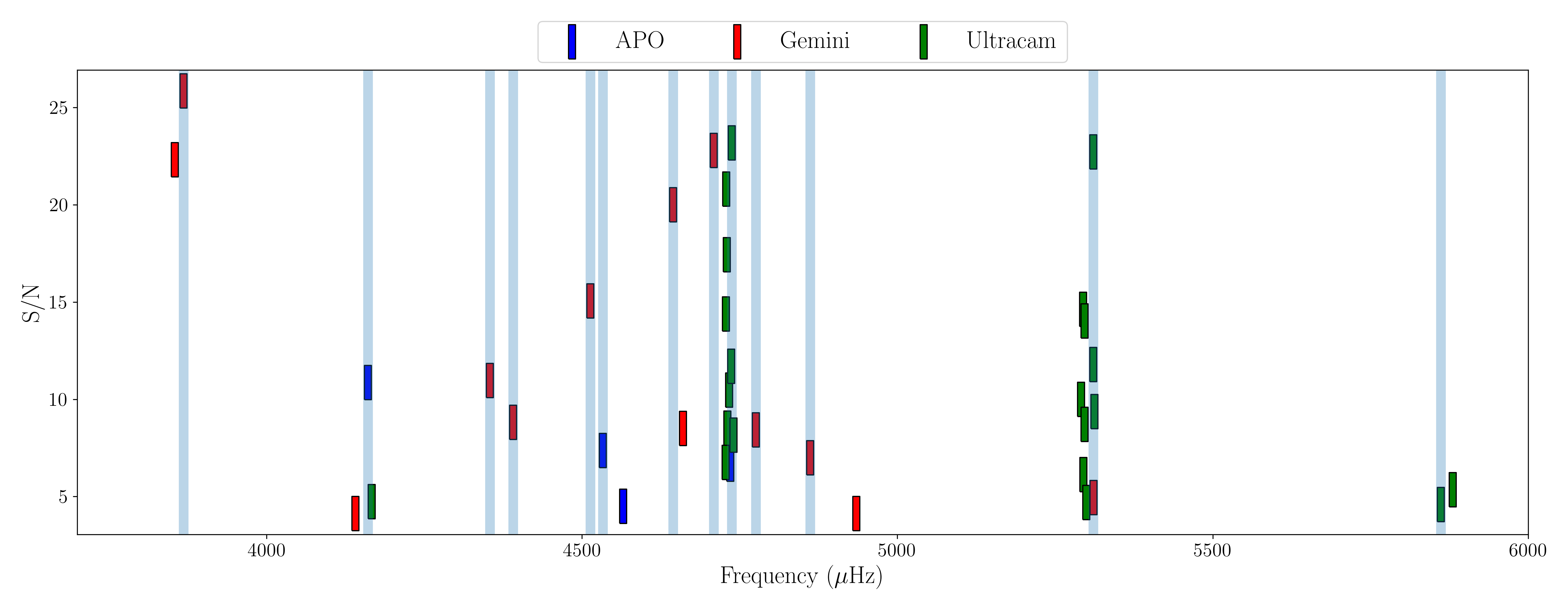} 
 \caption{Frequency distribution of WD J0049−2525, color-coded by instrument (ULTRACAM in green, Gemini in red, and APO in blue), showing frequency as a function of signal-to-noise ratio ($S/N$). The vertical lines represent the frequencies included in the seismic fit.}
    \label{fig:F_dist}
\end{figure*}

A prominent cluster of frequencies exists between 3900 $\mu$Hz and 6000 $\mu$Hz as can be seen in Figure \ref{fig:F_dist}. This cluster might represent closely spaced pulsation modes, potentially part of rotational multiplets, although no definitive pattern is evident. 
From the combined list, we identified three doublets where either the central component (\(m = 0\)) or one of the side components (\(m = -1\) or \(m = +1\)) is missing. The first doublet, 4390.8132 - 4353.9702 $\mu$Hz, has a frequency separation of 39 $\mu$Hz, corresponding to \(m = -1\) and \(m = +1\). The second, 4533.0211 - 4513.3539 $\mu$Hz, has a separation of 18 $\mu$Hz, representing \(m = -1\) and \(m = 0\). The third, 4775.7105 - 4737.4096 $\mu$Hz, has a separation of 38 $\mu$Hz, also corresponding to \(m = -1\) and \(m = +1\). From these candidates, we obtain an average frequency splitting of \(\langle \Delta\nu \rangle = 19.079\) $\mu$Hz, which corresponds to a rotation period of \(P_{\text{rot}} = 0.3\) days (7.28 hours).
Another weak candidate is located at 4728.7463 - 4737.4096 $\mu$Hz with a splitting of 8.66 $\mu$Hz, which would indicate a rotation period of 0.67 days (16.03 hours).
Both solutions are consistent with expectations for a high-mass WD \citep{Hermes17}. However, additional observations would be helpful for identifying the correct solution for WD J0049$-$2525's rotation period. 

\section{Asteroseismic modeling}
\label{model}

We have detected 13 pulsation modes and three combination frequencies in WD J0049$-$2525. 
Based on the Pan-STARRS $grizy$ photometry and {\it Gaia} parallax, \citep{kilic23merger} constrained the mass of this star to be $M = 1.26 \pm 0.01~M_\odot$ for an ONe core. 
This object presents one of the best chances to use asteroseismology to investigate the interior of a potential ONe-core WD to date.

\subsection{Period Spacing}
\label{psp}

\begin{figure}[htp]
    \centering
    \includegraphics[width=0.45\textwidth]{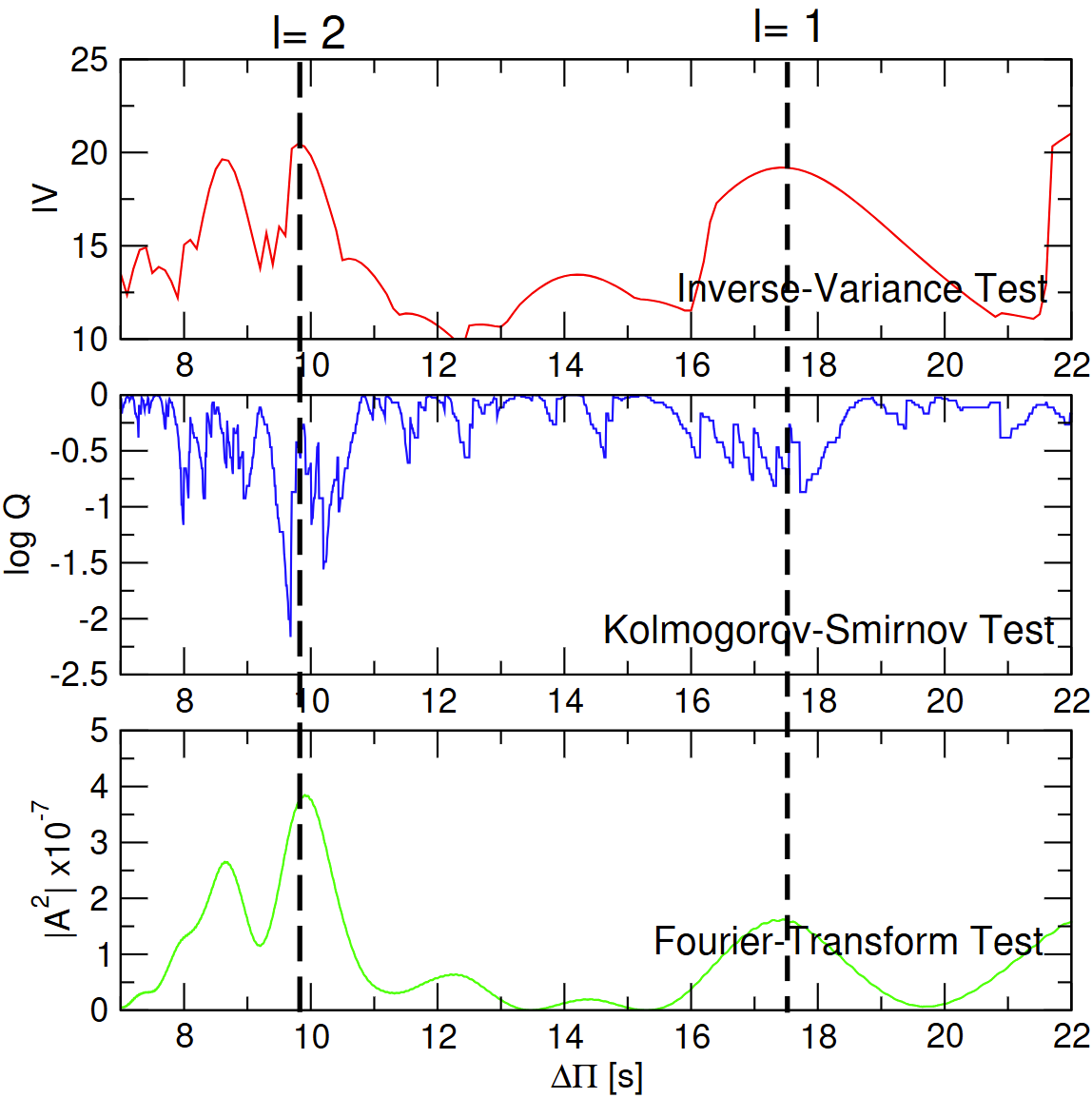}
    \caption{IV (upper panel), KS (middle panel), and FT (bottom panel)
significance tests to search for constant period spacings in the set of periods of WD J0049$-$2525. The vertical thick black dashed  lines indicate the possible period spacings present in the star as indicated by the tests (see text for details).}

    \label{fig:tests}
\end{figure}

To identify uniform period spacings ($\Delta \Pi$) within the period set of WD J0049$-$2525, we performed three statistical tests: Inverse Variance \citep[IV;][]{1994MNRAS.270..222O}, Kolmogorov-Smirnov \citep[KS;][]{1988IAUS..123..329K}, and Dirac comb with Fourier Transform \citep[FT;][]{1997MNRAS.286..303H}. Figure \ref{fig:tests} shows the results from this analysis. 

In the IV test, a peak in the inverse variance suggests a consistent period spacing. For the KS test, the quantity $Q$ represents the likelihood that the observed periods are randomly arranged. Therefore, any uniform or systematically non-random period spacing in the star's period spectrum will manifest as a minimum in $Q$. Lastly, in the FT test, we compute the Fourier Transform of a Dirac comb function (derived from the observed periods) and plot the square of the amplitude of the resulting function against the inverse of the frequency. A peak in the square of the amplitude indicates a constant period spacing. We observe two strong indications of period spacings at $17.56\pm1.02$ seconds and $9.79\pm0.52$ seconds (average values and uncertainties from the three tests), which can be linked to $\ell= 1$ and $\ell= 2$ modes, respectively. The theoretical relationship between dipole ($\ell= 1$) and quadrupole ($\ell= 2$) period spacings of $g$ modes according to asymptotic theory \citep{1980ApJS...43..469T} is $\Delta \Pi_{\ell= 2}= \Delta \Pi_{\ell= 1} / \sqrt{3}$. In this instance, the period spacings we find are in a ratio of $1.79$, which is close to $\sqrt{3}$. This suggests the presence of both $\ell= 1$ and $\ell= 2$ modes, indicating two distinct period spacings. 

\begin{figure}[htp]
    \centering
    \includegraphics[width=0.45\textwidth]{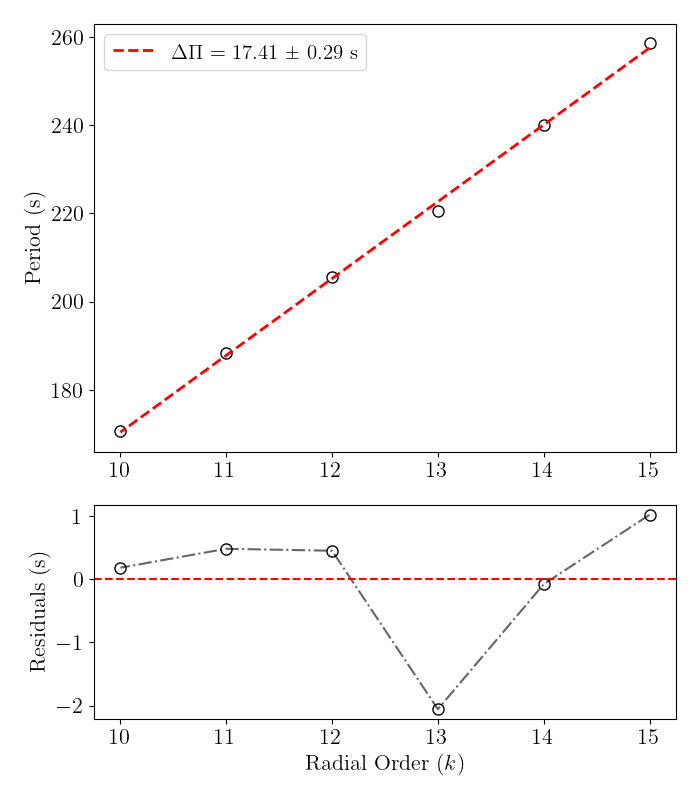}
    \caption{Upper panel: Least squares fit of the periods with a period spacing of $\approx 17$ sec, which correspond to modes with $\ell =1$. Lower panel: Residuals of the fit. The mode identified with $k=13$ is likely trapped.}
    \label{fig:l_identification}
\end{figure}

We identify a sequence of six modes that can be reliably classified as $\ell = 1$ modes with consecutive radial orders (see Table \ref{tab:l_identification} and Figure \ref{fig:l_identification}).  
Assigning the harmonic degree $\ell$ to these modes, we fit their periods as a function of radial order. We determine a period spacing of 17.41 seconds for $\ell = 1$.  
Using the theoretical ratio between $\ell = 1$ and $\ell = 2$ period spacings, we infer a period spacing of 10.04 seconds for $\ell = 2$, which agrees well with the results from three statistical tests,  
providing strong constraints on the period-to-period fits.

\begin{figure}[htp]
    \centering
    \includegraphics[width=0.45\textwidth]{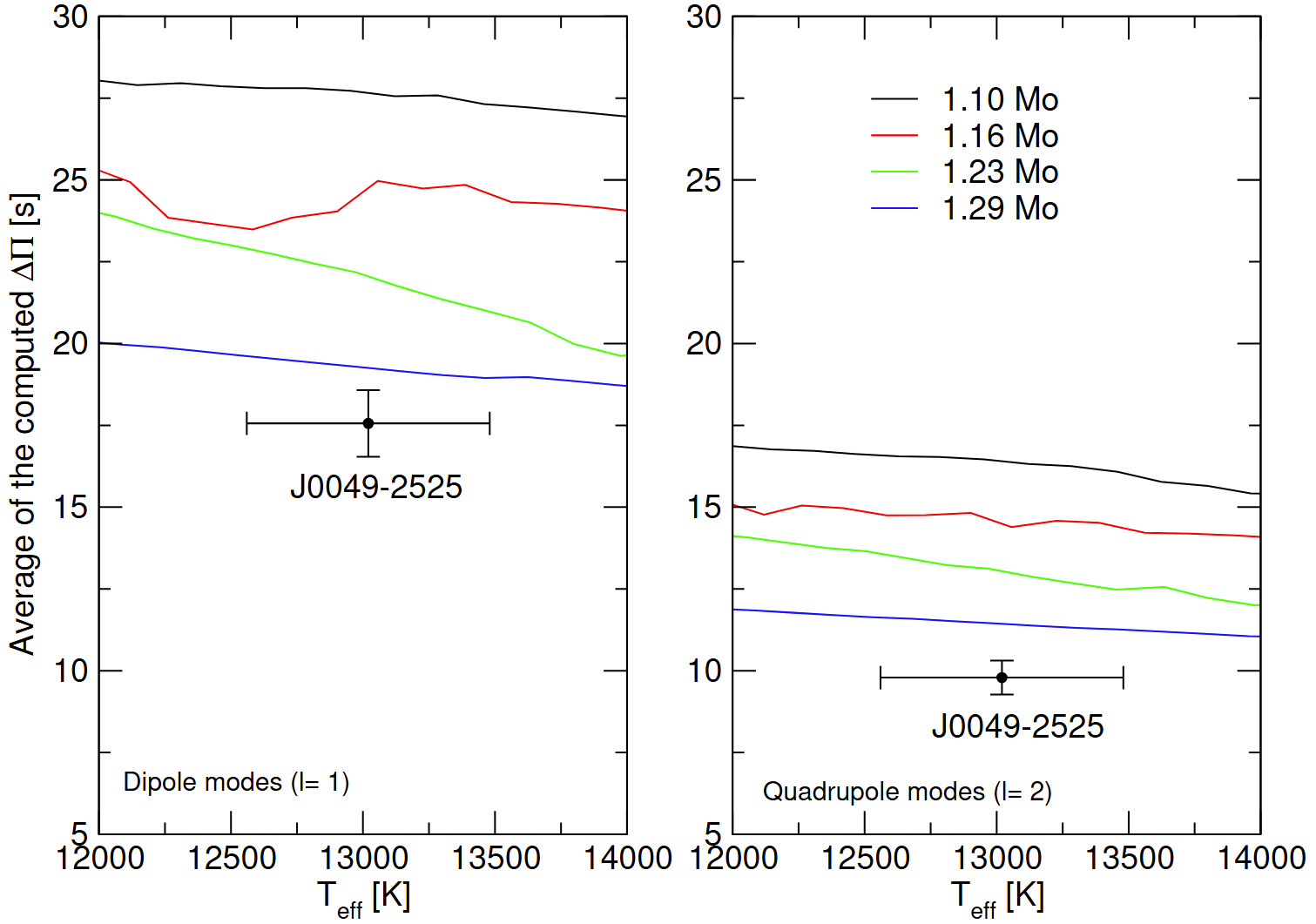}
    \caption{Average period spacings for $\ell= 1$ (left panel) and $\ell= 2$ (right panel) modes for ONe-core ultramassive
    WDs with thick H envelopes and masses ranging from 1.10 to $1.29~M_{\odot}$ (colored lines). The location of WD J0049$-$2525
    is based on the effective temperature derived by \cite{kilic23merger}, $T_{\rm eff}= 13\,020 \pm 460$ K, and the period spacings of $\Delta \Pi= 17.56 \pm1.02$ s for $\ell= 1$ and $\Delta \Pi= 9.79 \pm 0.52$ s for $\ell= 2$ modes.}
    \label{fig:ACPS}
\end{figure}

Assuming that the spacings of 17.56 seconds ($\ell= 1$) and 9.79 seconds ($\ell= 2$) are genuine, we can compare them with the average theoretical period spacings corresponding to various stellar masses at the star's effective temperature. This allows us to infer (or constrain) the stellar mass of WD J0049$-$2525. Here, we assume that the star harbours a core made of O and Ne, and employ the pulsation computations corresponding to the ONe-core ultra-massive WD evolutionary sequences employed in  \cite{Córsico19} and \cite{2019A&A...621A.100D}. Figure \ref{fig:ACPS} shows the results of this comparison. We conclude that WD J0049$-$2525 has a mass $M_{\star}\geq1.29\ M_{\odot}$. This finding aligns with the high mass suggested by spectroscopy (see Figure \ref{fig:spec}). 
Note that if one were to deny the existence of the 17-second signal and assume that the 10-second signal corresponds to
modes with $\ell=1$, this would imply a mass above the Chandrasekhar limit, which is impossible (see the left panel of Fig. \ref{fig:ACPS}).

\begin{table}
\setlength{\tabcolsep}{3pt}
\centering
\caption{Identification of a sequence of $\ell= 1$ modes with consecutive radial orders. The value of the radial order is tentative.}
\begin{tabular}{ccc}
\hline
\hline
$\Pi$ & $\ell$  & Trial $k$ \\
(s) &         &           \\
\hline
170.5994 &	1 & 10 \\
188.3111 &	1 & 11 \\
205.6936 &	1 & 12 \\
220.6000 &	1 & 13 \\
239.9992 &	1 & 14 \\
258.5046 &	1 & 15 \\
\hline
\hline
\end{tabular}%
\label{tab:l_identification}%
\end{table}%

\subsection{Period-to-Period Fits}
\label{ptpf}

In this section, we aim to find an evolutionary model whose theoretical periods  best match the individual pulsation periods detected for WD J0049$-$2525. The quality of the fit is computed by evaluating the quality function defined as follows:

\begin{equation}
\chi ^2(M_{\star},M_H,T_{\rm eff})=\frac{1}{N} \sum_{i=1}^{N} \rm min[(\Pi_i^{O}-\Pi_k^{th})^2],  
\end{equation}

\noindent where $N$ represents the number of detected modes, $\Pi_i^{\rm O}$ are the observed periods, and $\Pi_k^{\rm th}$ are the theoretically computed periods (where $k$ is the radial order). The best-fitting model is chosen by identifying the minimum value of $\chi^2$.

We use the same grid of ultramassive ONe-core WD models as in \cite{Kilic23}, that include evolutionary sequences of stellar masses $M_{\rm WD}/M_{\odot}= 1.10, 1.13, 1.16, 1.19, 1.22, 1.25, 1.29 $ and total H-content between $10^{-6}$ and $10^{-10}M_{\rm WD}$.
These evolutionary sequences were computed using the {\tt LPCODE} \citep{2005A&A...435..631A}, taking into account Coulombian diffusion \citep{2020A&A...644A..55A}. Further details are provided in \cite{2019A&A...625A..87C} and \cite{Córsico19}. The pulsational properties of our models were computed throughout the ZZ Ceti instability strip, employing the {\tt LP-PUL} pulsation code \citep{Córsico19}. We computed adiabatic pulsation periods of $\ell =1, 2$ g-modes in the range 70-1500 s, as is typically observed in ZZ Ceti stars.

The asteroseismic period-to-period fit analysis we performed considers different scenarios for the mode identification: a) none of the modes has an assigned $\ell$ value before the fit; b) 5 of the 6 modes shown in Table \ref{tab:l_identification} are assigned $\ell=1$, except for the 220 s mode, which shows the largest departure from the predicted value (see Figure~ \ref{fig:l_identification}).

\begin{table*}[]
\caption{Parameters of the best-fit models.}
  \centering
\begin{tabular}{cccccccccc}
\hline
\hline
Model & $M_{\star}$ & $\log g$ & $T_{\rm eff}$  & $\log\left(\frac{M_{\rm H}}{M_{\star}}\right)$ & $\frac{M_{\rm cryst}}{M_{\star}}$ & $\chi^2$& $BIC$ & $\ell= 1$ & $d$\\
\#      &   ($M_{\odot}$)   &  ($\rm cm~ s^{-2}$)     &  (K)   &          & (\%)       &     &  & & (pc)  \\

\hline
      &      &       &  $\ell$ free     &          &        &       & &&  \\
\hline
1     & 1.22 & 9.14      & 12\,794 & $-6.0$       &   96.34     & 5.61 & 1.00& 5 & 121.5 \\
2     & 1.25  & 9.25     &  13\,382     &  $-9.5$        &98.54        &  6.07  &  1.04  &4 & 113.6 \\

\hline
      &      &       &   5 $\ell=1$    &          &        &    &   & &  \\
\hline
3     & 1.29 &   9.38    & 13\,186 & $-7.5$     &   99.43     & 11.12&  1.30& 8& 98.1 \\
4     & 1.29 &  9.38     & 12\,539 & $-8.5$     &   99.66     & 10.72 & 1.28&8& 93.3\\
5     & 1.29 &  9.39     & 13\,065 & $-9.0$     &   99.59     & 12.59 &  1.35& 8  &96.2  \\

\hline
\hline
\end{tabular}
\label{tab:best-fits}
\end{table*}

In Table \ref{tab:best-fits} we tabulated the stellar parameters for the best-fit models for both scenarios. When $\ell$ is left as a free parameter, the periods are mostly fitted with the $\ell=2$ modes. This is expected due to the smallest period spacing for the $\ell=2$ modes. We find solutions with effective temperatures around $13\, 000$ K and with a wide variety of H-content, ranging from thick envelopes ($10^{-6}M_{\star}$) to thin envelopes ($10^{-9.5}M_{\star}$). However, the asteroseismic distance estimated for these solutions is significantly different from the {\it Gaia} distance of $99.7^{+2.9}_{-2.7}$ pc \citep{Bailer-Jones21}. Hence, they are ruled out based on {\it Gaia} astrometry.

Upon pre-assigning the five $\ell=1$ modes before the period-to-period fit, the asteroseismic solutions are more massive ($1.29 M_{\odot}$) and in better agreement with the photometric and spectroscopic determinations of $T_{\rm eff}$ and $\log{g}$. All of these solutions are characterized by a low H content ($\lesssim 10^{-7.5}M_{\rm WD}$) and a crystallized portion of the ONe core of $>99$\%. 
Finally, the distance estimated for our best-fit models is in the range $[93.3-98.1]$ pc, showing a much better agreement with the {\it Gaia} distance than when $\ell$ is left as a free parameter.

It is important to note that, given the proximity between some of the observed modes (see Table \ref{tab:Frequencies}), whose proximity could be due to them being modes with unresolved rotation or magnetic multiplets (same $\ell$ but different $m$) or even two close $\ell=1$ and $\ell=2$ modes, those nearby modes are fitted with the same theoretical mode. To explore the impact of these closely spaced modes on our analysis, we performed an additional asteroseismic fit to the period list in which we selected the central values of the F4, F5, F6 and F8, F9, F10 modes. With this, we fitted the following periods: 170.59, 188.31, 205.69, 211.08, 215.29, 220.60, 229.67, 239.99, 258.50 s. We still find the same best-fitting solutions as in \#4 and \#5 in Table \ref{tab:best-fits}, with a slight shift in temperature, with $T_{\rm eff}= 12\,897$ and $13\,496$ K, respectively. 
In order to have an indicator of the quality of the period fit, we computed the Bayes Information Criterion \citep[$BIC$;][]{2000MNRAS.311..636K}:

\begin{equation}
   BIC=N_p \left(\frac{\log N}{N}\right) + \log(\sigma^2),
\end{equation}
\noindent where $N_p$ is the number of free parameters in the models and $N$ is the number of observed periods. The smaller the value of $BIC$, the better the quality of the fit. In our case, $N_p = 3$ (stellar mass, effective temperature, and thickness of the H envelope), and $N = 13$. We list the $BIC$ values together with the best fit models' parameters in Table \ref{tab:best-fits}. All of the $BIC$ values are between $\sim 1$ and $\sim 1.35$, meaning that all of these fits are good. For comparison, \cite{Corsico21} obtained $BIC= 0.59$, 1.15, and 1.20 for the Planetary Nebula Nucleus Variable (PNNV) stars RX J2117+3412, NGC 1501, and NGC 2371, respectively, and $BIC = 1.18$ for the hybrid DOV star HS $2324+3944$. Similarly, \cite{Bischoff-Kim19} and \cite{Corsico22}  obtained $BIC = 1.20$ and 1.13, respectively, for the prototypical DBV WD GD~358.

We note that our asteroseismic analysis is limited by the fact that the stellar mass derived is at the edge of our model grid, so an extension of this grid to larger masses would be worthwhile. It would also be desirable to repeat our period-to-period asteroseismic analyses with CO-core ultramassive WD models for comparison. The extension of the grid of models to higher masses and the computation of a new grid of CO-core ultramassive WD models is currently in progress (De Gerónimo et al 2025), and will be used in the future to test the core composition of pulsating ultramassive WDs, including J0049$-$2525.
\newpage
\section{Summary and Conclusions}
\label{conclusion}

We present a detailed observational and asteroseismic analysis of the most massive pulsating WD currently known, WD J0049$-$2525, based on time-series photometry from three different telescopes. 
Our frequency analysis reveals a rich spectrum of pulsation modes, with several prominent frequencies concentrated in the range between 3868 $\mu$Hz (258 s) and 5861 (170 s) $\mu$Hz. The combined data set from the three observatories enabled us to detect 13 significant pulsation frequencies, many of which have high signal-to-noise ratios. We identified two potential frequency splittings, indicating a rotation period of either 0.3 d (7.28 h) or 0.67 d (16.03 h). The former (0.3 d) is a stronger candidate, but both are
in agreement with expectations for such a massive WD.

We use three different statistical tests to search for uniform period spacings, and find strong evidence for consistent spacings at 17.56 s and 9.79 s that can be linked to $\ell= 1$ and $\ell= 2$ modes, respectively. The ratio between these two spacings is remarkably close to the expected ratio of $\sqrt{3}$ between the dipole and quadruple period spacings of g-modes according to asymptotic theory. Comparison of these observed period spacings with those calculated for different stellar masses and effective temperatures allows us to rule out stellar masses below $\sim$1.29 $M_{\sun}$ for WD J0049$-$2525.

Detailed asteroseimic period-to-period fits analysis using ONe-core models reveals that the best-fitting models are characterized by a stellar mass of 1.29$M_{\odot}$, with a thin H-envelope $\lesssim 10^{-7.5}M_{\rm WD}$ and
a crystallized core mass fraction of $>99\%$. The derived asteroseismic distance of 93.3-98.1 pc is in excellent agreement with
the {\it Gaia} inferred distance of $99.7^{+2.9}_{-2.7}$ pc \citep{Bailer-Jones21}. 

\citet{corsico23} investigated the impact of General Relativity (GR) on g-mode pulsations in ultramassive WDs, and
demonstrated that the resulting pulsation periods can be up to 50\% shorter (for the most massive WDs with $M=1.369~M_{\odot}$), when a relativistic treatment is used. However, they also demonstrated that the GR effects on the g-mode periods of WD J0049$-$2525 are smaller than 1\%. Hence, WD J0049$-$2525 is not massive enough for the exploration of the GR effects on WD pulsations.

In conclusion, the combination of photometric observations and appropriate asteroseismic models opens up a new avenue for the study of the interiors of ultramassive WDs. WD J0049$-$2525, with its complex pulsation spectrum, offers a valuable opportunity to test theoretical models of stellar evolution and interior physics. However, further high-precision observations and a more refined modeling approach are necessary to fully understand the detailed structure of this star, especially to constrain its rotation period and to verify the presence of additional subtle pulsation modes that might provide deeper insights into the physics of ultramassive WDs.

\section{Acknowledgments}

We appreciate the detailed report from the anonymous referee and their valuable suggestions that improved the content of the article. \"{O}\c{C} thanks TÜBİTAK (The Scientific and Technological Research Council of Türkiye) for funding this research under the 2214-A research project at the University of Oklahoma, as part of his doctoral thesis. MU gratefully acknowledges funding from the Research Foundation Flanders (FWO) by means of a junior postdoctoral fellowship (grant agreement No. 1247624N). IP acknowledges support from a Royal Society University Research Fellowship (URF\textbackslash R1\textbackslash 231496). 

This work is supported in part by the NSF under grant AST-2205736, the NASA under grants 80NSSC22K0479, 80NSSC24K0380,
and 80NSSC24K0436, the NSERC Canada, and the Fund FRQ−NT (Québec).
This work was supported by PIP 112-200801-00940 grant from CONICET, grant G149 from the University of La Plata, PIP-2971 from CONICET (Argentina) and by PICT 2020-03316 from Agencia I+D+i (Argentina). This work was partially supported by the MINECO grant PID2023-148661NB-I00 and by the AGAUR/Generalitat de Catalunya grant SGR-386/2021.

Based on observations obtained with the Apache Point Observatory 3.5-meter telescope, which is owned and operated by the Astrophysical Research Consortium.

Based on observations collected at the European Organisation for Astronomical Research in the Southern Hemisphere under ESO programme 0112.D-0386(A).

Based on observations obtained at the international Gemini Observatory, a program of NSF NOIRLab, which is managed by the Association of Universities for Research in Astronomy (AURA) under a cooperative agreement with the U.S. National Science Foundation on behalf of the Gemini Observatory partnership: the U.S. National Science Foundation (United States), National Research Council (Canada), Agencia Nacional de Investigaci\'{o}n y Desarrollo (Chile), Ministerio de Ciencia, Tecnolog\'{i}a e Innovaci\'{o}n (Argentina), Minist\'{e}rio da Ci\^{e}ncia, Tecnologia, Inova\c{c}\~{o}es e Comunica\c{c}\~{o}es (Brazil), and Korea Astronomy and Space Science Institute (Republic of Korea).

VSD and ULTRACAM are funded by the Science and Technology Facilities Council (grant ST/Z000033/1).

\software{Period04
\citep{Lenz05}, {\tt LPCODE} \citep{2005A&A...435..631A}, {\tt LP-PUL} \citep{2006A&A...454..863C}}

\facilities{ARC 3.5m (ARCTIC), NTT (ULTRACAM), Gemini:South (GMOS spectrograph)}

\bibliography{pulsator}
\bibliographystyle{aasjournal}

\appendix
\section{Appendix A}
\label{sec:Appendix}

\begin{figure*}[htp]
    \centering
    \includegraphics[width=1.0\textwidth]{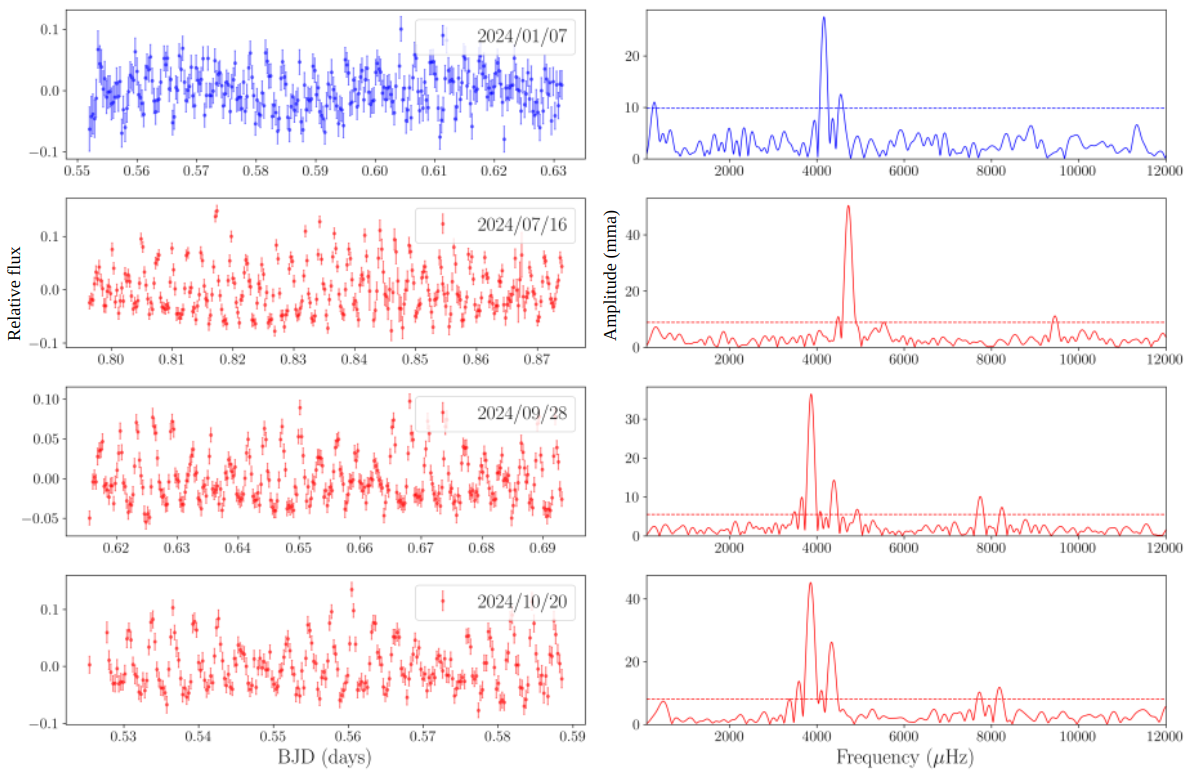}
    \caption{Light curves (left) and corresponding Fourier transforms (right) for APO (blue) and Gemini (red) observations of WD J0049$-$2525. Dashed horizontal lines indicate the 4×$S/N$ detection threshold for each band. The observations are sorted chronologically and detailed in Table 2.}
    \label{fig:APO-GEMINI}
\end{figure*}

The light curves (left) and Fourier transforms (right) of WD J0049-2525 observed within the scope of the study are shown in Figure \ref{fig:APO-GEMINI}. The light curve and FT (blue) in the upper panel of the figure were obtained from the observation at APO on 2024-01-07. The light curves and FTs in the lower three panels (red) were obtained from the observations at the Gemini Observatory on 2024-07-16, 2024-09-28, and 2024-10-20.

The frequency solutions obtained from all observations used in this work are presented in Table \ref{tab:Full_FT}. The top panel shows the frequencies obtained from the observations made at APO. The second panel shows the frequencies obtained from Gemini observations. The bottom three panels show the frequencies obtained from the ULTRACAM data in the $ugr$ filters. The lines with an asterisk (*) in the dates represent frequencies based on the data from \citet{Kilic23}.
\newpage

\begin{table*}
\scriptsize
\setlength{\tabcolsep}{8pt}
\centering
\caption{Frequency solution for all observations.}
\begin{tabular}{l r r r l}
\hline
\hline
\multicolumn{1}{c}{$\nu$} & \multicolumn{1}{c}{$\Pi$} & \multicolumn{1}{c}{$A$} & \multicolumn{1}{c}{$S/N$} & \multicolumn{1}{c}{Date} \\
\multicolumn{1}{c}{($\mu$Hz)} & \multicolumn{1}{c}{(s)} & \multicolumn{1}{c}{(mma)} &  & \multicolumn{1}{c}{(yyyy.mm.dd)}  \\
\hline
\hline
\multicolumn{5}{c}{APO 3.5 m} \\
$4160.7 \pm 6.3$  & $240.3 \pm 0.4$ & $27.2 \pm 2.1$ & $10.86$ & 2024-01-07 \\
$4533.0 \pm 8.3$  & $220.6 \pm 0.4$ & $28.0 \pm 3.0$ & $7.36$  & 2022-12-22$^{*}$ \\
$4565.3 \pm 15.1$ & $219.0 \pm 0.7$ & $11.2 \pm 2.1$ & $4.49$  & 2024-01-07 \\
$4735.3 \pm 13.4$ & $211.2 \pm 0.6$ & $32.0 \pm 4.2$ & $6.66$  & 2023-01-08$^{*}$ \\
\hline
\multicolumn{5}{c}{Gemini} \\
$3854.7 \pm 3.6$  & $259.4 \pm 0.2$ & $44.6 \pm 1.6$ & $22.31$ & 2024-10-20 \\
$3868.4 \pm 2.3$  & $258.5 \pm 0.2$ & $36.2 \pm 1.0$ & $25.85$ & 2024-09-28 \\
$4141.0 \pm 19.4$ & $241.5 \pm 1.1$ & $8.3 \pm 1.6$  & $4.13$  & 2024-10-20 \\
$4354.0 \pm 7.3$  & $229.7 \pm 0.4$ & $21.9 \pm 1.6$ & $10.96$ & 2024-10-20 \\
$4390.9 \pm 6.6$  & $227.7 \pm 0.3$ & $12.4 \pm 1.0$ & $8.82$  & 2024-09-28 \\
$4513.4 \pm 5.8$  & $221.6 \pm 0.3$ & $25.6 \pm 1.3$ & $15.05$ & 2022-12-27$^{*}$ \\
$4644.7 \pm 0.4$  & $215.3 \pm 0.1$ & $40.0 \pm 1.4$ & $20.00$ & 2022-12-26$^{*}$ \\
$4660.4 \pm 0.4$  & $214.6 \pm 0.1$ & $17.0 \pm 1.4$ & $8.50$  & 2022-12-26$^{*}$ \\
$4709.1 \pm 3.0$  & $212.4 \pm 0.1$ & $50.1 \pm 1.8$ & $22.79$ & 2024-07-16 \\
$4775.7 \pm 10.3$ & $209.4 \pm 0.5$ & $14.3 \pm 1.3$ & $8.42$  & 2022-12-27$^{*}$ \\
$4861.6 \pm 9.8$  & $205.7 \pm 0.4$ & $15.4 \pm 1.8$ & $6.99$  & 2024-07-16 \\
$4935.1 \pm 14.1$ & $202.6 \pm 0.6$ & $5.8 \pm 1.0$  & $4.13$  & 2024-09-28 \\
$5311.0 \pm 17.5$ & $188.3 \pm 0.6$ & $8.4 \pm 1.3$  & $4.95$  & 2022-12-27$^{*}$ \\
$7729.3 \pm 18.6$ & $129.4 \pm 0.3$ & $8.6 \pm 1.6$  & $4.29$  & 2024-10-20 \\
$7751.9 \pm 8.6$  & $129.0 \pm 0.1$ & $9.5 \pm 1.0$  & $6.77$  & 2024-09-28 \\
$8189.2 \pm 15.1$ & $122.1 \pm 0.2$ & $10.6 \pm 1.6$ & $5.30$  & 2024-10-20 \\
$8232.3 \pm 12.4$ & $121.5 \pm 0.2$ & $6.6 \pm 1.0$  & $4.72$  & 2024-09-28 \\
$9218.5 \pm 23.3$ & $108.5 \pm 0.3$ & $75.7 \pm 1.4$ & $5.50$  & 2022-12-26$^{*}$ \\
$9291.4 \pm 15.3$ & $107.6 \pm 0.2$ & $9.6 \pm 1.3$  & $5.66$  & 2022-12-27$^{*}$ \\
$9459.0 \pm 13.8$ & $105.7 \pm 0.2$ & $10.9 \pm 1.8$ & $4.97$  & 2024-07-16 \\
\hline
\hline
\multicolumn{5}{c}{ULTRACAM $u$-band} \\
$4728.0 \pm 2.3$ & $211.5 \pm 0.1$ & $33.1 \pm 4.5$ & $6.75$ & 2023-10-06 \\
$4730.5 \pm 1.7$ & $211.4 \pm 0.1$ & $3.8 \pm 3.7$  & $8.52$ & 2023-10-05 \\
$4740.1 \pm 1.7$ & $211.0 \pm 0.1$ & $31.0 \pm 3.2$ & $8.16$ & 2023-10-07 \\
$5295.1 \pm 2.4$ & $188.9 \pm 0.1$ & $27.0 \pm 3.7$ & $6.13$ & 2023-10-05 \\
$5299.6 \pm 3.3$ & $188.7 \pm 0.1$ & $23.0 \pm 4.5$ & $4.69$ & 2023-10-06 \\
$5312.3 \pm 1.5$ & $188.2 \pm 0.1$ & $35.6 \pm 3.2$ & $9.36$ & 2023-10-07 \\
\hline
\multicolumn{5}{c}{ULTRACAM $g$-band} \\
$576.0 \pm 2.1$  & $1736.2 \pm 6.3$ & $8.0 \pm 1.0$  & $6.69$  & 2023-10-07 \\
$4166.7 \pm 3.2$ & $240.0 \pm 0.2$  & $7.6 \pm 1.4$  & $4.74$  & 2023-10-05 \\
$4728.7 \pm 0.7$ & $211.5 \pm 0.1$  & $33.3 \pm 1.4$ & $20.81$ & 2023-10-05 \\
$4729.8 \pm 0.8$ & $211.4 \pm 0.1$  & $31.4 \pm 1.5$ & $17.43$ & 2023-10-06 \\
$4731.6 \pm 0.1$ & $211.3 \pm 0.1$  & $29.4 \pm 0.8$ & $32.64$ & 2023-10-05/06/7 \\
$5294.4 \pm 1.0$ & $188.9 \pm 0.1$  & $23.4 \pm 1.4$ & $14.62$ & 2023-10-05 \\
$5296.9 \pm 1.0$ & $188.8 \pm 0.1$  & $25.2 \pm 1.5$ & $14.03$ & 2023-10-06 \\
$5310.4 \pm 0.6$ & $188.3 \pm 0.1$  & $27.3 \pm 1.0$ & $22.72$ & 2023-10-07 \\
$5870.1 \pm 0.4$ & $170.4 \pm 0.1$  & $5.2 \pm 0.8$  & $5.81$  & 2023-10-05/06/07 \\
$5880.3 \pm 2.6$ & $170.1 \pm 0.1$  & $6.4 \pm 1.0$  & $5.35$  & 2023-10-07 \\
$9480.9 \pm 3.4$ & $105.5 \pm 0.1$  & $5.0 \pm 1.0$  & $4.16$  & 2023-10-07 \\
$10022.8 \pm 3.7$& $99.8 \pm 0.1$   & $6.6 \pm 1.4$  & $4.15$  & 2023-10-05 \\

\hline
\multicolumn{5}{c}{ULTRACAM $r$-band} \\
$4728.4 \pm 1.1$ & $211.5 \pm 0.1$ & $33.1 \pm 4.5$ & $14.38$ & 2023-10-05 \\
$4733.3 \pm 1.4$ & $211.3 \pm 0.1$ & $27.2 \pm 2.3$ & $10.47$ & 2023-10-06 \\
$4736.8 \pm 1.2$ & $211.1 \pm 0.1$ & $21.1 \pm 1.5$ & $11.70$ & 2023-10-07 \\
$5291.1 \pm 1.5$ & $189.0 \pm 0.1$ & $23.0 \pm 4.5$ & $9.99$  & 2023-10-05 \\
$5297.0 \pm 1.7$ & $188.8 \pm 0.1$ & $22.6 \pm 2.3$ & $8.70$  & 2023-10-06 \\
$5297.7 \pm 0.1$ & $188.8 \pm 0.1$ & $17.7 \pm 0.8$ & $19.68$ & 2023-10-05/06/07 \\

\hline
\hline
\end{tabular}
\label{tab:Full_FT}
\end{table*}

\end{document}